\documentclass[aps,prb,twocolumn,amsmath,amssymb,groupedaddressx,longbibliography
]{revtex4-2}

\usepackage{graphicx}
\usepackage{dcolumn}
\usepackage{bm}
\usepackage{multirow}
\usepackage{braket}
\usepackage{color}
\usepackage{ulem}
\usepackage{times}
\usepackage{booktabs}

\begin{document}
\title{
Essential role of anisotropic magnetic dipole in anomalous Hall effect
}

\author{Satoru Hayami$^1$ and Hiroaki Kusunose$^2$}
\affiliation{
$^1$Department of Applied Physics, the University of Tokyo, Tokyo 113-8656, Japan \\
$^2$Department of Physics, Meiji University, Kawasaki 214-8571, Japan 
 }

\begin{abstract}
We theoretically investigate the anomalous Hall effect (AHE) that requires neither a net magnetization nor an external magnetic field in collinear antiferromagnets. 
We show that such an emergent AHE is essentially caused by a ferroic ordering of the anisotropic magnetic dipole (AMD), which provides an effective coupling between ordered magnetic moments and electronic motion in the crystal. 
We demonstrate that the AMD is naturally induced by the antiferromagnetic ordering, in which the magnetic moments have a quadrupole spatial distribution. 
In view of the ferroic AMD ordering, we analyze the behavior of the AHE in the orthorhombic lattice system, where the AHE is largely enhanced by the large coupling between the AMD and the spin-orbit interaction. 
From these findings, the AMD can be used as a descriptor in general to investigate the ferromagnetic-related physical quantities in antiferromagnets including noncollinear ones, which is detectable by using the x-ray magneto-circular dichroism. 
\end{abstract}

\maketitle

Anomalous Hall effect (AHE) is one of the fundamental physical responses in condensed matter physics~\cite{hall1881xviii,Smith_PhysRev.17.23,Nagaosa_RevModPhys.82.1539}. 
Although the AHE was originally discussed in ferromagnets with the spin-orbit interaction~\cite{Karplus_PhysRev.95.1154,smit1958spontaneous,Maranzana_PhysRev.160.421,Berger_PhysRevB.2.4559,nozieres1973simple,Jungwirth_PhysRevLett.88.207208,Gosalbez_PhysRevB.92.085138}, it is an up-to-date subject as it has also been observed in the noncollinear~\cite{Tomizawa_PhysRevB.80.100401,Chen_PhysRevLett.112.017205,nakatsuji2015large,Suzuki_PhysRevB.95.094406,Chen_PhysRevB.101.104418} and noncoplanar antiferromagnets~\cite{Ohgushi_PhysRevB.62.R6065,Shindou_PhysRevLett.87.116801,taguchi2001spin,Neubauer_PhysRevLett.102.186602} even with a negligibly small magnetization. 
In any cases, the Berry curvature generated from the effective hoppings on a closed loop plays a role of an emergent magnetic field in momentum space, which is the origin of the AHE~\cite{Ye_PhysRevLett.83.3737,Haldane_PhysRevLett.93.206602,Xiao_RevModPhys.82.1959,Zhang_PhysRevB.101.024420}.

Recently, other types of the AHE have theoretically been suggested in collinear antiferromagnets, which are referred to as the crystal Hall effect~\cite{vsmejkal2020crystal,feng2020observation,shao2020interfacial,samanta2020crystal}.  
The crystal Hall effect is caused by the symmetry lowering by the arrangement of the nonmagnetic atoms in collaboration with time-reversal symmetry breaking~\cite{vsmejkal2020crystal}. 
Moreover, such an AHE in collinear antiferromagnets emerges even without the nonmagnetic ions when the magnetic ions break the mirror symmetry as the uniform magnetization does. 
Such phenomena were demonstrated in La$M$O$_3$ ($M=$ Cr, Mn, and Fe)~\cite{Solovyev_PhysRevB.55.8060}, bilayer MnPSe$_3$~\cite{Sivadas_PhysRevLett.117.267203}, NiF$_2$~\cite{li2019quantum}, and in the organic antiferromagnet by the authors and their collaborators~\cite{Naka_PhysRevB.102.075112}. 

In the present study, we aim at clarifying that it is an anisotropic magnetic dipole (AMD) that the key ingredient is for the AHE. 
The AMD has the same rotational property as the ordinary magnetic dipoles, i.e., the spin and orbital angular momenta, but it carries {\it no} net magnetic moment. 
This is in a family of electronic multipoles, which is clearly distinguished from the other types of 
multipoles, such as the magnetic octupoles~\cite{Kusunose_JPSJ.77.064710,kuramoto2009multipole,Santini_RevModPhys.81.807}, magnetic toroidal dipoles~\cite{Fiebig0022-3727-38-8-R01,Spaldin_0953-8984-20-43-434203,Hayami_PhysRevB.90.024432,hayami2018microscopic}, and so on. 
The AMD is an observed quantity that is defined as $\bm{M}'=[3(\bm{r}\cdot\bm{\sigma})\bm{r}-r^{2}\bm{\sigma}]/\sqrt{10}$ with $\bm{r}$ and $\bm{\sigma}$ being the position vector within the atomic wavefunction and spin~\cite{kusunose2020complete}, which is often called as $\bm{T}$ vector in the context of the x-ray magneto-circular dichroism (XMCD)~\cite{Carra_PhysRevLett.70.694,Stohr_PhysRevLett.75.3748,stohr1995x,crocombette1996importance}. 
We demonstrate the essential role of the AMD in the AHE by considering the collinear antiferromagnetic ordering in the orthorhombic lattice system as the simplest example. 
Since the concept of the AMD is universal irrespective of crystal structures and types of magnetic orderings, it will enhance insight to engineer magnetic materials showing the large AHE. 
We also propose that the AMD is useful as a descriptor for the ferromagnetic-related physical quantities in antiferromagnets including noncollinear ones such as Mn$_3$Sn~\cite{yamasaki2020augmented}.

Let us start by introducing the property of the AMD ($\bm{M}'$). 
As was already mentioned, $\bm{M'}$ has the same rotational property besides the spatial-inversion and time-reversal properties as the magnetic dipoles, namely, it is the time-reversal odd rank-1 axial tensor (pseudovector)~\cite{Kusunose_JPSJ.77.064710,kuramoto2009multipole,Santini_RevModPhys.81.807}. 
In contrast to the fact that the ordinary magnetic dipoles are the fundamental spin pseudovector, $\bm{\sigma}$, or the contraction of two polar vectors, $-i(\bm{r}\times\bm{p})$, the AMD is constructed by the contraction of the rank-2 polar tensors and $\bm{\sigma}$. 
Indeed, its components are explicitly written as 
\begin{align}
\label{eq:AMDx}
M'_x&=\sqrt{\frac{3}{10}}\left[ \left(-\frac{1}{\sqrt{3}}Q_u + Q_v\right)\sigma_x+Q_{xy} \sigma_y + Q_{zx} \sigma_z\right], \\
\label{eq:AMDy}
M'_y &= \sqrt{\frac{3}{10}}\left[Q_{xy}\sigma_x -\left(\frac{1}{\sqrt{3}}Q_u + Q_v \right)\sigma_y+Q_{yz} \sigma_z \right], \\
\label{eq:AMDz}
M'_z &=\sqrt{\frac{3}{10}}\left[Q_{zx} \sigma_x + Q_{yz}\sigma_y+\frac{2}{\sqrt{3}}Q_u \sigma_z\right],
\end{align}
where $(Q_u, Q_v, Q_{yz}, Q_{zx}, Q_{xy})$ are electric quadrupoles with the functional forms of $[(3z^2-r^2)/2, \sqrt{3}(x^2-y^2)/2, \sqrt{3}yz, \sqrt{3}zx, \sqrt{3}xy]$~\cite{kusunose2020complete}. 

\begin{figure}[t!]
\begin{center}
\includegraphics[width=1.0 \hsize]{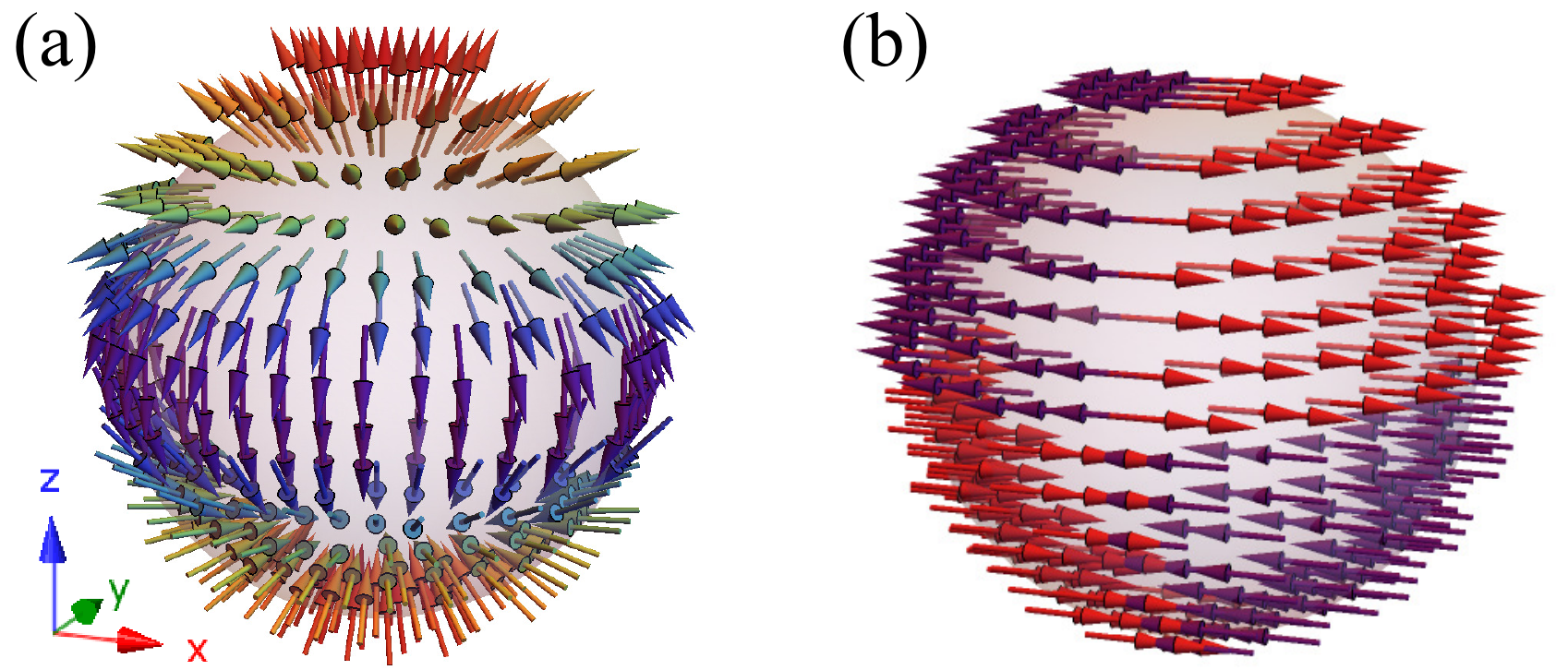} 
\caption{
\label{Fig:ponti}
Schematic pictures of the spin polarization of the anisotropic magnetic dipole (a) $M'_z$ and (b) its $\sigma_x$ component in real space [$\bm{r}=(x,y,z)$]. 
In (a) and (b), the arrows represent the spin polarization at each $\bm{r}$ and the colors denote the spatial distribution of $\sigma_z$ and $\sigma_x$ components ($Q_{u}$ and $Q_{zx}$) in Eq.~(\ref{eq:AMDz}).
}
\end{center}
\end{figure}

The expressions of $\bm{M}'$ in Eqs.~(\ref{eq:AMDx})-(\ref{eq:AMDz}) clearly show that the AMD is accompanied with a spatially anisotropic distribution of $\bm{\sigma}$. 
Figure~\ref{Fig:ponti}(a) shows a schematic picture of the spin polarization of $M'_z$ at each $\bm{r}=(x,y,z)$ on the sphere. 
The $x$-, $y$- and $z$-spin components in Eq.~(\ref{eq:AMDz}) show symmetric quadrupole distributions as $zx$, $yz$, and $3z^2-r^2$, respectively, as exemplified for $\sigma_x$ in Fig.~\ref{Fig:ponti}(b). 
The angle dependence of $\bm{M}'$, which is obtained by substituting the polar coordinates both in the spin and quadrupole components, is the same as that of the ordinary magnetic dipoles, i.e., $\bm{M}' \propto (\sin \theta \cos \phi,\sin\theta \sin\phi, \cos \theta)$. 
The same anisotropy of $\bm{M}'$ results in the same symmetry structure in physical responses such as the AHE, magneto-optical Kerr effect, and Nernst effect. 
In other words, the concept of the AMD is naturally applicable to the phenomenological linear-response tensor, in which the electric monopole, quadrupole, and magnetic dipole appear in the conductivity tensor for instance~\cite{Hayami_PhysRevB.98.165110}. 

Nevertheless, $\bm{M}'$ does not carry any magnetic moment due to the anisotropic spatial distribution of spins. 
This means that there is no direct coupling between $\bm{M}'$ and an external magnetic field, and hence it does not appear in multipole expansions of the scalar and vector potentials~\cite{kusunose2020complete}. 
It is noteworthy that $\bm{M}'$ is independent from the higher-rank magnetic multipoles and magnetic toroidal multipoles as well as the magnetic dipoles in continuous rotational symmetry~\cite{hayami2018microscopic,Hayami_PhysRevB.98.165110,kusunose2020complete}. 
Indeed, all the multipoles including the AMD in Eqs.~(\ref{eq:AMDx})-(\ref{eq:AMDz}) satisfy the mutual orthogonality. 
From these properties, the AMD could be a hidden degree of freedom that characterizes the ferromagnetic-related physics even without a net magnetization. 
Moreover, this aspect has great advantage to an efficient spintronics devices without the leakage of magnetic field as ordinary ferromagnetism does~\cite{Baltz_RevModPhys.90.015005}.

\begin{figure}[t!]
\begin{center}
\includegraphics[width=1.0 \hsize]{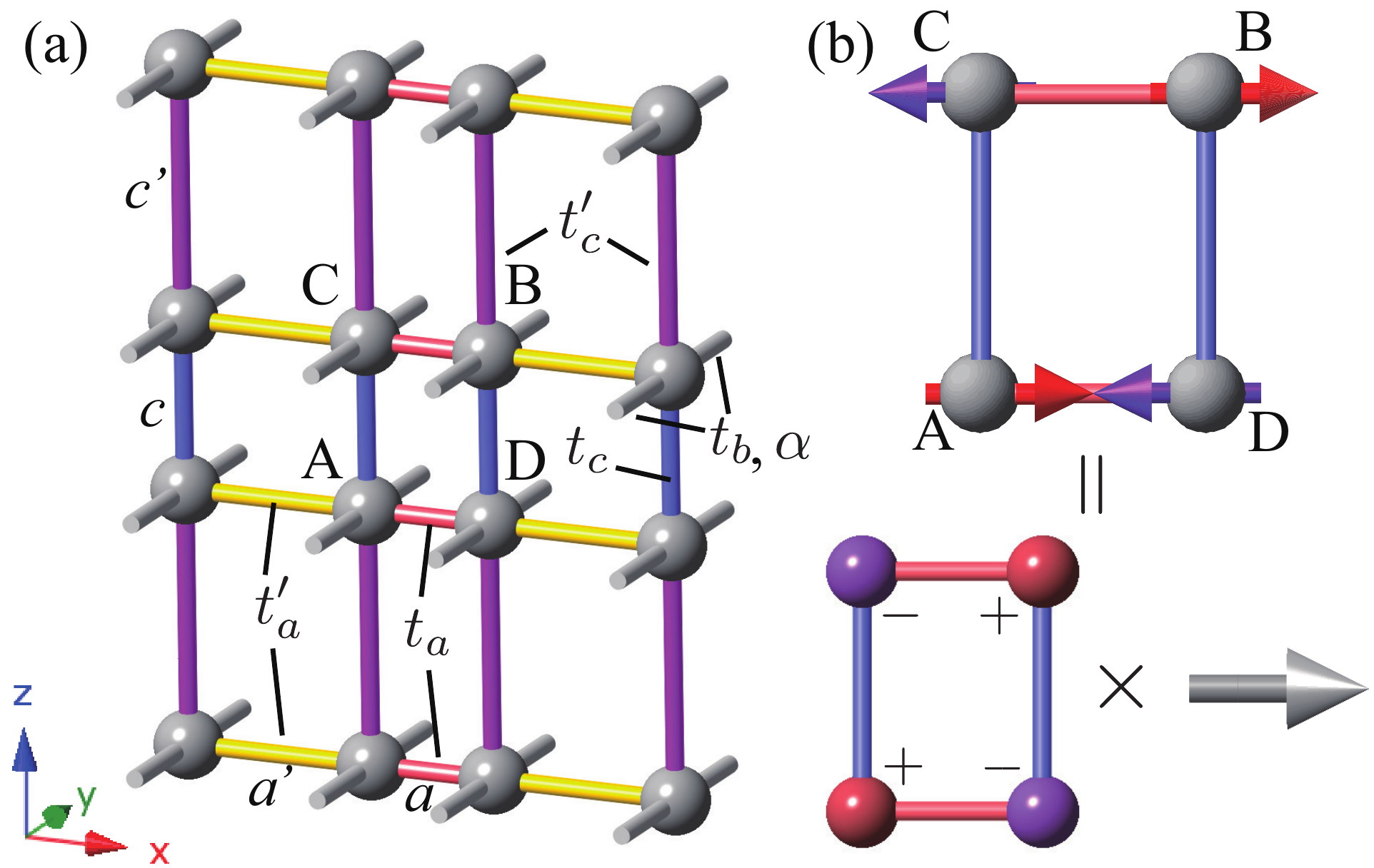} 
\caption{
\label{Fig:Lattice}
(a) Orthorhombic lattice structure consisting of four sublattices (A-D) with hopping parameters ($t_a, t_b, t_c, t'_a, t'_c, \alpha$). 
(b) Collinear antiferromagnetic structure, which can be decomposed into the cluster electric quadrupole $Q_{zx}$ and spin $\sigma_{x}$. 
}
\end{center}
\end{figure}

With these preliminaries, we elucidate the importance of the AMD on the basis of a specific lattice system in the collinear antiferromagnetic state that shows the AHE. 
Figure~\ref{Fig:Lattice}(a) shows our minimal model in the orthorhombic four-sublattice (A-D) structure under the space group $D_{2\rm h}^{1}$.
The orbital degrees of freedom are not taken into account for simplicity. 
The system consists of two alternating bonds along the $x$ and $z$ directions in the $xz$ plane with the lattice constants $a+a'$ and $c+c'$, respectively, and the $xz$ plane is stacked along the $y$ direction with the lattice constant $b$. 
We take the lattice constant as $a=a'=b=c=c'=1$ for notational simplicity and the difference between $a$ and $a'$ (or $c$ and $c'$) is expressed as the different hopping amplitudes. 
The model Hamiltonian is given by 
\begin{align}
\label{eq:Model_Ham}
\mathcal{H}=& \sum_{ij\sigma} (t_{ij} c^{\dagger}_{i\sigma}c_{j\sigma}  + {\rm h.c.})+ 2  \sum_{l \bm{k} \sigma \sigma'}\alpha_l \sin k_y   c^{\dagger}_{l\bm{k}\sigma} \sigma_x^{\sigma\sigma'}c_{l\bm{k}\sigma'} \nonumber \\
&- \sum_{l \bm{k} \sigma\sigma'} h_l c^{\dagger}_{l\bm{k}\sigma}\sigma_x^{\sigma\sigma'}c_{l\bm{k}\sigma'}
\equiv\mathcal{H}_{t}+\mathcal{H}_{\rm SOC}+\mathcal{H}_{\rm MF}, 
\end{align}
where $c^{\dagger}_{i \sigma}$ ($c_{i\sigma}$) is the creation (annihilation) operator of the electron at site $i$ with spin $\sigma$. 
$c^{\dagger}_{l \bm{k} \sigma}$ is the Fourier transform of $c^{\dagger}_{i \sigma}$ where $l$ denotes the sublattice index A-D. 
The first term in Eq.~(\ref{eq:Model_Ham}) represents the hopping between $i$th and $j$th sites. 
We consider five hopping parameters on the different bonds; $t_a$ and $t'_a$ along the $x$ direction, $t_b$ along the $y$ direction, and $t_c$ and $t'_c$ along the $z$ direction, as shown in Fig.~\ref{Fig:Lattice}(a).  
The hopping parameters are set as $t_a=1$, $t'_a=0.5$, $t_b=0.7$, $t_c=0.4$, and $t'_c=0.2$. 
The second term represents the site-dependent spin-orbit interaction. 
Although similar hoppings like the $y$-spin component with the $k_{x}$ dependence is also allowed by symmetry, we take into account only the $x$-spin component with the $k_y$ dependence, since it is essential to obtain the AHE.
As this term originates from the alternating stacking along the $z$ direction as well as the atomic spin-orbit coupling, the sign of $\alpha_l$ depends on the sublattice, $-\alpha_{\rm A}=\alpha_{\rm B}=\alpha_{\rm C}=-\alpha_{\rm D}\equiv \alpha$. 
Note that $\alpha_{l}\sin k_{y}$ and $\sigma_{x}$ belong to the same irreducible representation (irrep.), $B_{3g}^{-}$. 
The third term represents the mean-field term corresponding to the magnetic order. 
Among the twelve possible magnetic ordering patterns within the four sublattice, we suppose the collinear antiferromagnetic order with $h_{\rm A}=h_{\rm B}=-h_{\rm C}=-h_{\rm D}\equiv h$, which only exhibits the AHE without the net magnetization, as discussed below. 

Figure~\ref{Fig:Lattice}(b) represents the supposed collinear antiferromagnetic structure. 
The magnetic moments in the sublattices A and B (C and D) are along the positive (negative) $x$ direction with the equal amplitude. 
This collinear order is regarded as a product of the anisotropic alignment of point charges and the spin along the $x$ direction.
Since the charge distribution is characterized by the $zx$-type quadrupole, the third term in Eq.~(\ref{eq:Model_Ham}) is proportional to $Q_{zx}\sigma_x$, which is nothing but the first term of $M'_z$ in Eq.~(\ref{eq:AMDz}) when we adopt the concept of the cluster multipole by replacing $\bm{r}$ in Eq.~(\ref{eq:AMDz}) with the position vectors of the sublattices from the center of the plaquette ADBC~\cite{Suzuki_PhysRevB.99.174407,Hayami_PhysRevB.102.144441}. 
It is noted that, in general, the antiferromagnetic structure corresponding to the cluster AMD can be obtained by the virtual cluster method, in which ambiguity such as the choice of the origin is eliminated~\cite{Suzuki_PhysRevB.99.174407}.
In the end, this collinear antiferromagnetic ordering is equivalent to the ferroic order of the cluster AMD.
Therefore, we refer this antiferromagnetic ordering to as the (ferroic cluster) AMD ordering. 
As shown below, it is the key ingredient that gives rise to the AHE.
Since $Q_{zx}$ ($h_{l}$) and $\sigma_{x}$ belong to $B_{2g}^{+}$ and $B_{3g}^{-}$, respectively, $\mathcal{H}_{\rm MF}$ represents the symmetry-breaking term belonging to $B_{1g}^{-}$, which is the same irrep. of $M_{z}'$.
The irreps. of the related quantities of the present paper in $D_{\rm 2h}$ ($C_{2h}$) point group are summarized in Table~\ref{tab_irrep}. 

\begin{table}[t!]
\caption{
Irreducible representations of the related quantities in this paper in $D_{\rm 2h}$ ($C_{2h}$) point group. 
}
\label{tab_irrep}
\centering
\begin{tabular}{llcll}
\hline\hline
irrep. & quantities & $\,\,\,$ & irrep. & quantities \\
\hline 
$A_{g}$ ($A_{g}$) & $Q_{u}$, $Q_{v}$, $\mathcal{H}_{t}$, $\mathcal{H}_{\rm SOC}$ & & $A_{u}$ ($A_{u}$) & --- \\
$B_{1g}$ ($B_{g}$) & $M_{z}'$, $\sigma_{z}$, $Q_{xy}$, $\phi$, $\mathcal{H}_{\rm MF}$ & & $B_{1u}$ ($B_{u}$) & $\sin k_{z}$, $\alpha_{l}$  \\
$B_{2g}$ ($A_{g}$) & $M_{y}'$, $\sigma_{y}$, $Q_{zx}$, $h_{l}$
& & $B_{2u}$ ($A_{u}$) & $\sin k_{y}$ \\
$B_{3g}$ ($B_{g}$) & $M_{x}'$, $\sigma_{x}$, $Q_{yz}$ & & $B_{3u}$ ($B_{u}$) & $\sin k_{x}$ \\
\hline \hline
\end{tabular}
\end{table}

\begin{figure}[t!]
\begin{center}
\includegraphics[width=1.0 \hsize]{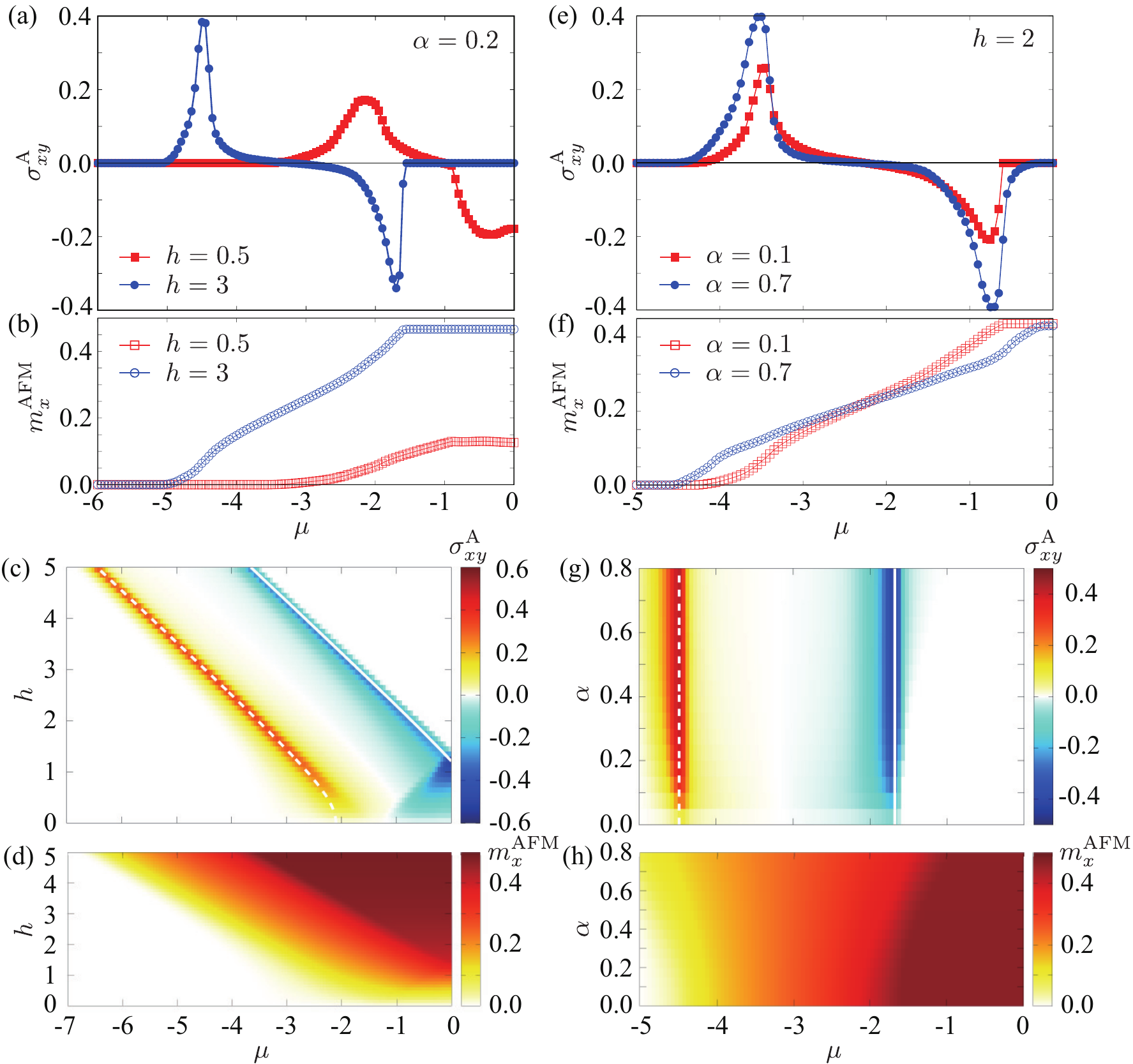} 
\caption{
\label{Fig:Hall}
(a) $\mu$ dependences of $\sigma^{\rm A}_{xy}$ with fixed $\alpha=0.2$ for $h=0.5$ (red square) and $h=3$ (blue circle)
(b) Staggered magnetization $m^{\rm AFM}_{x}$ corresponding to (a). 
(c, d) Contour plots of (c) $\sigma^{\rm A}_{xy}$ and (d) $m^{\rm AFM}_{x}$ in the plane of $\mu$ and $h$ at $\alpha=0.2$. 
(e, f) The same plot as (a, b) with fixed $h=2$ for $\alpha=0.1$ (red square) and $\alpha=0.7$ (blue circle). 
(g, h) The same plot as (c, d) in the plane of $\mu$ and $\alpha$ at $h=2$. 
In (c) and (g), the dashed and solid lines represent the eigenvalues at R and U points in the Brillouin zone as shown in the inset of Fig.~\ref{Fig:dos}(a). 
}
\end{center}
\end{figure}

The Hall conductivity, $\sigma^{\rm A}_{xy}\equiv\sigma_{xy}=-\sigma_{yx}$, in the AMD ordered state is calculated by using the Kubo formula with scattering rate $\tau^{-1}=10^{-3}$ and the temperature $T=10^{-4}$. 
The summation of the momentum $\bm{k}$ is taken over $120^3$ grid points in the first Brillouin zone. 
Figure~\ref{Fig:Hall}(a) shows $\sigma^{\rm A}_{xy}$ as a function of the chemical potential $\mu$ for $h=0.5$ and $3$ at $\alpha=0.2$. 
The results for $\mu>0$ is omitted owing to the particle-hole symmetry in the model in Eq.~(\ref{eq:Model_Ham}). 
The staggered magnetization $m^{\rm AFM}_x\equiv\braket{\sigma_{x}}$, which is equivalent to the expectation value of the ferroic cluster AMD, is also shown in Fig.~\ref{Fig:Hall}(b). 
As shown in Figs.~\ref{Fig:Hall}(a) and \ref{Fig:Hall}(b), $\sigma^{\rm A}_{xy}$ becomes finite in the AMD ordered state $m_x^{\rm AFM} \ne 0 $ except for the insulating region for $|\mu| \lesssim 1.55$ for $h=3$. 
This result means that the AHE is induced by the metallic AMD ordering. 
Nevertheless, the behavior of $\sigma^{\rm A}_{xy}$ and $m_x^{\rm AFM}$ against $\mu$ appears to be totally different. 
For both $h$, $\sigma^{\rm A}_{xy}$ exhibits the two peak structures while varying $\mu$: $\mu \simeq -2.15$ and $\mu \simeq -0.35$ for $h=0.5$ and $\mu \simeq -4.5$ and $\mu \simeq -1.7$ for $h=3$. 
Meanwhile, $m_x^{\rm AFM}$ shows a monotonous increase for both $h$. 
The two peaks in $\sigma^{\rm A}_{xy}$ move to smaller $\mu$ and their values are enhanced with increase of $h$. 
These behaviors are common for other $h$, as shown in Figs.~\ref{Fig:Hall}(c) and \ref{Fig:Hall}(d). 

Moreover, as shown in Fig.~\ref{Fig:Hall}(e) with fixed $h=2$, the peak positions of $\sigma^{\rm A}_{xy}$ are almost unchanged while varying $\alpha$, but the peak values are enhanced with increase of $\alpha$. 
As similar to the results in Figs.~\ref{Fig:Hall}(a) and \ref{Fig:Hall}(b), the behavior of $\sigma^{\rm A}_{xy}$ is qualitatively different from that of $m_x^{\rm AFM}$. 
The overall behaviors of $\sigma^{\rm A}_{xy}$ and $m_x^{\rm AFM}$ against $\alpha$ are shown in Figs.~\ref{Fig:Hall}(g) and \ref{Fig:Hall}(h), which indicates that there is almost no $\alpha$ dependence of the peak positions. 
It is noted that $\sigma^{\rm A}_{xy}$ vanishes at $\alpha=0$.

\begin{figure}[t!]
\begin{center}
\includegraphics[width=1.0 \hsize]{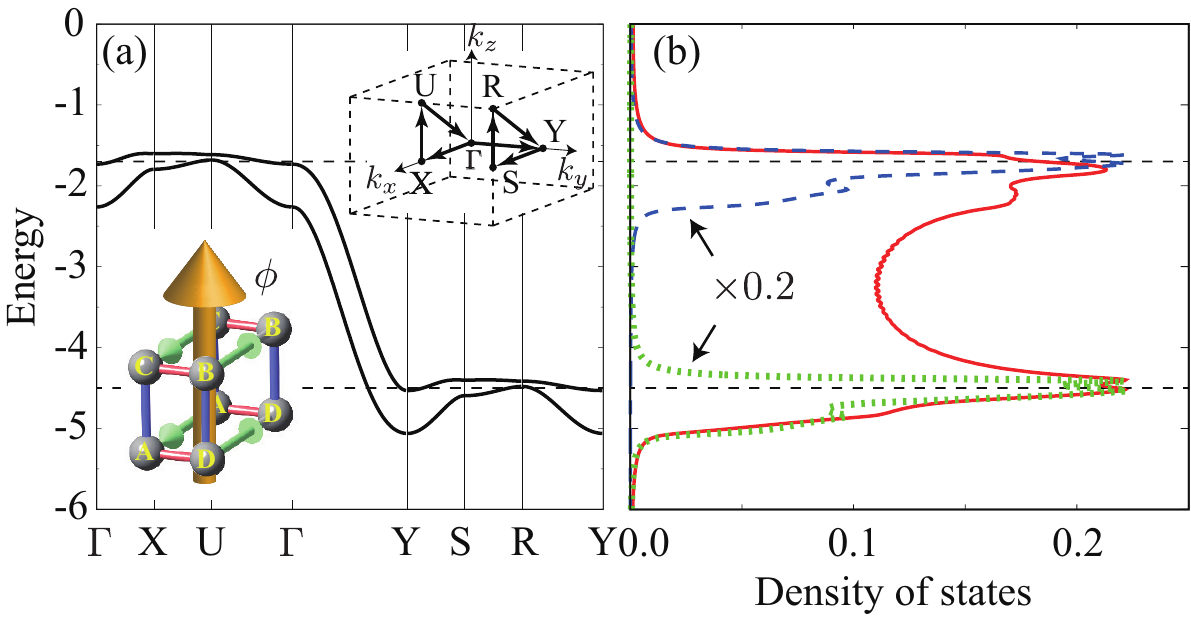} 
\caption{
\label{Fig:dos}
(a) Electronic band structure along the symmetry lines of the Brillouin zone in the right-top inset at $\alpha=0.2$ and $h=3$. 
The left-bottom inset indicates the schematic picture of an effective flux $\phi$ (thick orange arrow) and the effective imaginary hopping (green arrows on the bond) under the AMD. 
(b) Density of states (DOS) with the same parameters as (a). 
The red curves stand for the total DOS, while the blue dashed (green dotted) curves represent the DOS at $k_y=0$ ($k_y=\pi$). 
}
\end{center}
\end{figure}

From the above results, it is concluded that the AMD ordering is necessary to induce $\sigma_{xy}^{\rm A}$ but their behaviors do not have simple correlation like $\sigma_{xy}^{\rm A}\propto m_{x}^{\rm AFM}$ except in the vanishing limit of $m_{x}^{\rm AFM}$.
Hence, let us elucidate the essential terms for the AHE as follows. 
At first, $\sigma^{\rm A}_{xy}$ arises from the nonzero Berry curvature in the presence of $m_x^{\rm AFM}$. 
In other words, the electrons feel the effective magnetic flux when they move in the closed loop in real space~\cite{Zhang_PhysRevB.101.024420,Naka_PhysRevB.102.075112}.
Then, in order to deduce the effective magnetic flux, let us introduce an orbital angular-momentum operator on the plaquette ADDA as shown in the inset of Fig.~\ref{Fig:dos}(a) as 
$\mathcal{O}=\mathcal{O}_{y}+\mathcal{O}_{x}$ with $\mathcal{O}_{y}= \sin k_y \sum_{l\sigma} \gamma_l\, c^{\dagger}_{l\sigma}c_{l\sigma}$ ($-\gamma_{\rm A}=\gamma_{\rm B}=-\gamma_{\rm C}=\gamma_{\rm D}=1$) and $\mathcal{O}_{x}\propto \sin k_{x}$.
By evaluating the quantity, $\rho ={\rm Tr}[e^{-\beta \mathcal{H}_{\bm{k}}} \mathcal{O}]$, where $\mathcal{H}=\sum_{\bm{k}}\mathcal{H}_{\bm{k}}$ and $\beta=1/T$, we obtain nonzero $\rho$ as $\rho 
\propto \alpha h \sin^2 k_y$ in the lowest order for $\beta \mathcal{H}_{\bm{k}} \ll 1$~\cite{Hayami_PhysRevB.101.220403,Hayami_PhysRevB.102.144441}. 
Note that $\mathcal{O}_{x}$ does not contribute to $\rho$ in the present model. 
This indicates that the effective magnetic flux coupled with $\mathcal{O}$ under the AMD ordering is represented by $\alpha_l h_l\sin k_y $. 
In the real-space picture, this magnetic flux $\phi$ on the plaquette in the $xy$ plane is depicted in the inset of Fig.~\ref{Fig:dos}(a). 
On the neighboring plaquettes within the unit cell, the fluxes penetrate in opposite direction, however, the summation over these fluxes remains finite due to the inequivalent plaquettes, leading to the net magnetic flux over the crystal. 
From the symmetry viewpoint, the effective magnetic flux along the $z$ direction belongs to the irrep. $B^-_{1g}$ in the present case. 
One can confirm that from Table.~\ref{tab_irrep}, $\alpha_l h_l\sin k_y$ actually belongs to $B^-_{1g}$ as it does. 

The above result suggests that an important coupling between the AMD order parameter and the site-dependent spin-orbit interaction can be identified by the symmetry argument in collinear antiferromagnets in general.
The form factor $f_{\alpha}(\bm{k})$ and spin operator $\sigma_{\alpha}$ belong to the same irrep. in $\mathcal{H}_{\rm SOC}$, while the product of irreps. of $\sigma_{\alpha}$ and the cluster quadrupole $Q_{\beta}$ belongs to the same irrep. of the AMD, $M'_{\zeta}$ in $\mathcal{H}_{\rm MF}$.
Therefore, the coupling terms so as to satisfy $f_{\alpha}(\bm{k})\otimes Q_{\beta}\in M'_{\zeta}$ are essential.
In the noncollinear case, two- or three-spin components are simultaneously involved in the couplings, but the analysis is straightforward.

The enhancement of $\sigma^{\rm A}_{xy}$ at particular $\mu$ in Fig.~\ref{Fig:Hall} is understood from the electronic state based on the effective emergent magnetic flux $\alpha_l h_l \sin k_y$. 
$\sigma^{\rm A}_{xy}$ is obtained from the current-current correlation function, and the relevant electronic current operator arising from the emergent magnetic flux contains the $\bm{k}$-derivative of it, i.e., $\alpha_l h_l\cos k_y$. 
This means that $\sigma^{\rm A}_{xy}$ is expected to be large for $k_y=0$ or $k_y = \pi$. 
In the present model parameters, we find that the electronic band dispersions in the $k_x$-$k_z$ plane at $k_y=0$ or $k_y = \pi$ are considerably flat in Fig.~\ref{Fig:dos}(a), which give rise to the large density of states at the peak positions of $\sigma^{\rm A}_{xy}$ in Fig.~\ref{Fig:dos}(b) in the case of $\alpha=0.2$ and $h=3$. 
In particular, the small energy difference between the bands in the $k_x$-$k_z$ plane at $k_y=0$ and $k_y=\pi$ is important to the large $\sigma^{\rm A}_{xy}$, where the energy difference is approximately expressed as $2|(t_a-t'_a)(t_c-t'_c)/h|$ in the limit of $h \gg t_a, t'_a, t_c, t'_c$.  
Indeed, the peak positions are well fitted by the eigenvalues at $R$ and $U$ points [the inset of Fig.~\ref{Fig:dos}(a)] where the energy difference becomes the smallest, as shown in the dashed and solid lines in Figs.~\ref{Fig:Hall}(c) and \ref{Fig:Hall}(g), respectively. 
Thus, the large enhancement of $\sigma^{\rm A}_{xy}$ is ascribed to the flat-band-like electronic structure separated by the small energy difference in the $\bm{k}$ points, in which the current arising from the effective magnetic flux, $\alpha_l h_l\cos k_y,$ is maximized. 
The opposite sign of $\sigma^{\rm A}_{xy}$ at the two peaks is attributed to the opposite sign of $\cos k_y$  in the expression of the effective current at $R$ and $U$ points.

\begin{table}[t!]
\caption{
Nonzero anomalous Hall conductivity ($\sigma_{\mu\nu}=-\sigma_{\nu\mu}\equiv \sigma^{\rm A}_{\mu\nu}$ for $\mu, \nu=x,y,z$) under the AMD ordering, $\bm{m}^{\rm AFM} \ne 0$ in the orthorhombic and monoclinic systems. 
In the orthorhombic systems, the necessary quadrupole component $Q_{\mu\nu}$ is shown, while in the monoclinic systems, the additional $Q_{\mu\nu}$ to that of the orthorhombic systems is shown. 
Note that $Q_{u}$ and $Q_{v}$ are omitted as they belong to the totally symmetric irrep. 
}
\label{tab_multipoles2}
\centering
\begin{tabular}{cccccccccc}
\hline\hline
$\bm{M}'$ & Hall tensor & &   \multicolumn{3}{c}{orthorhombic} & &  \multicolumn{3}{c}{monoclinic} \\
& $\bm{m}^{\rm AFM}\parallel$ & $\,\,\,$ &  $\bm{\hat{x}}$ & $\bm{\hat{y}}$ & $\bm{\hat{z}}$ & $\,\,\,$ & $\bm{\hat{x}}$ & $\bm{\hat{y}}$ & $\bm{\hat{z}}$ \\
\hline 
$M'_x$ & $\sigma_{yz}^{\rm A}$ & & --- & $Q_{xy}$ & $Q_{zx}$ & & $Q_{zx}$ & $Q_{yz}$ & ---  \\
$M'_y$ & $\sigma_{zx}^{\rm A}$ & & $Q_{xy}$ & --- & $Q_{yz}$ & & $Q_{yz}$ & $Q_{zx}$ & $Q_{xy}$ \\
$M'_z$ & $\sigma_{xy}^{\rm A}$ & & $Q_{zx}$ & $Q_{yz}$ & --- & & --- & $Q_{xy}$ & $Q_{zx}$ \\
\hline \hline
\end{tabular}
\end{table}

It is emphasized that the concept of the AMD can be applied to other systems with different symmetry, since the symmetry of the AMD is always the same as that of the ordinary magnetic dipoles, as discussed above. 
Therefore, we can predict a nonzero AME by investigating whether the electric quadrupole degree of freedom coupled to spin is active or not in terms of the atomic, cluster, and bond degrees of freedom of electrons in the framework of the augmented multipole description~\cite{Hayami_PhysRevB.98.165110,Hayami_PhysRevB.102.144441}. 
A similar argument is straightforwardly applied to the other high-symmetry lattice systems, such as cubic, tetragonal, hexagonal, and trigonal lattice systems. 
It is noted that in the monoclinic lattice system, additional coupling between $Q_{\mu\nu}$ and $\bm{\sigma}$ can appear since $\sigma_x$ and $\sigma_z$ belong to the same irrep. with the principal axis $y$. 
For example, the finite Hall response $\sigma^{\rm A}_{xy}$ is expected when the coupling $c_1 Q_{zx}\sigma_x+(c_2 Q_{yz}+c'_2 Q_{xy})\sigma_y +c_3' Q_{zx} \sigma_z$ becomes active. 
The recent theoretical finding of the AHE in the $\kappa$-type organic antiferromagnet is caused by such an additional coupling ($c_{i}'\ne0$)~\cite{Naka_PhysRevB.102.075112}. 
The necessary couplings for the appearance of the AHE in the orthorhombic and monoclinic lattice systems are summarized in Table~\ref{tab_multipoles2}.

Since the AMD is indistinguishable from the ordinary magnetic dipoles in the sense of symmetry, it is useful to introduce the AMD as a descriptor of the ferromagnetic-related physical phenomena in antiferromagnets with negligibly small magnetization. 
Note that when the size of the relevant Hilbert space is sufficiently large, the AMD is independent of the magnetic octupoole with the same irrep. where their matrices are different.
In the present work, we mainly discuss the cluster extension of the AMD, however, it also induces an atomic-scale AMD in the same irrep. 
For example, the noncollinear antiferromagnetic ordering in Mn$_3$Sn showing the large AHE~\cite{nakatsuji2015large,Suzuki_PhysRevB.95.094406} is also regarded as the ferroic cluster AMD ordering as similar to the present work, which should accompany the atomic-scale AMD in the same irrep. 
Indeed, it has been observed by means of XMCD measurement~\cite{yamasaki2020augmented}.

In summary, we have theoretically investigated the role of the AMD with particular emphasis on the AHE. 
To this end, we have generalized the concept of the atomic-scale AMD in the context of XMCD to the cluster one in order to characterize the ferromagnetic-related physics without the net magnetization under the antiferromagnetic ordering. 
The AMD provides a guiding principle to attain the large AHE in terms of not only the symmetry but also the microscopic parameters in Hamiltonian. 
Indeed, we have shown that the AHE is largely enhanced when the effective coupling between the AMD ordered parameter and the spin-orbit interaction is maximized by analyzing the fundamental orthorhombic four-sublattice system. 
As the concept of the AMD can be used generally for any noncollinear and noncoplanar magnetic structures irrespective of the space group symmetry, it is a natural descriptor for a further exploration of the materials with exhibiting the large AHE and Nernst effect in antiferromagnetic spintronics.

\begin{acknowledgments}
The authors thank Y. Yanagi, M. Naka, and R. Oiwa for fruitful discussions. 
This research was supported by JSPS KAKENHI Grants Numbers JP15H05885, JP18H04296 (J-Physics), JP18K13488, JP19K03752, JP19H01834, and by JST PREST (JPMJPR20L8). 
Parts of the numerical calculations were performed in the supercomputing systems in ISSP, the University of Tokyo, and in the MAterial science Supercomputing system for Advanced MUlti-scale simulations towards NExt-generation-Institute for Materials Research (MASAMUNE-IMR) of the Center for Computational Materials Science, Institute for Materials Research, Tohoku University.
\end{acknowledgments}

\bibliographystyle{apsrev}
\bibliography{ref}

\end{document}